\documentclass[reprint,amsfonts,amsmath,amssymb,aps,pra,twocolumn]{revtex4-2}
\usepackage[utf8]{inputenc}
\usepackage{graphicx}
\usepackage{physics}
\usepackage{dcolumn}
\usepackage{bm}
\usepackage{color}
\usepackage{ulem}

\begin{document}

\title{Evidence of ac-Stark-shifted resonances in intense two-color circularly polarized laser fields}
\author{Philipp Stammer}
\email{stammer@mbi-berlin.de}
\affiliation{Max-Born-Institute, Max-Born-Strasse 2A, Berlin D-12489, Germany}
\author{Serguei Patchkovskii}
\affiliation{Max-Born-Institute, Max-Born-Strasse 2A, Berlin D-12489, Germany}
\author{Felipe Morales}
\affiliation{Max-Born-Institute, Max-Born-Strasse 2A, Berlin D-12489, Germany}

\date{\today}

\begin{abstract}
We report on the appearance of a structure at low energies in the photo-electron momentum distribution of the hydrogen atom exposed to two-color counter-rotating bi-circular laser fields. These structures, which arise due to AC-Stark shifted resonances, break the three-fold symmetry, typical for the $\omega-2\omega$ bi-circular fields. We discuss the physical origin of this structure in terms of partial-wave interference between direct ionization channels and a resonant pathway, that passes through the AC-Stark shifted state and show how the underlying Rydberg state population depend on the field strength and pulse duration. 
\end{abstract}

\maketitle

\section{\label{sec:intro}Introduction}

When an atom is exposed to an electric field, the energy eigenstates of the atom are shifted due to additional interaction terms in the Hamiltonian, and the resulting phenomena include the Autler-Townes effect \cite{autler1955stark} leading to the observation of the Mollow-Triplet in resonant fluorescence, cooling and trapping of atoms in optical lattices, and many other related phenomena \cite{mollow1975, dalibard1989laser, cohen1998atom}.\\
In the limit of high laser frequencies $\omega \gg E_i$ , where $E_i$ is the binding energy of the $i$-th atomic eigenstate, the dynamical Stark shift is given by the cycle averaged kinetic energy of an electron in the laser field, i.e. the ponderomotive energy $U_p$, and level splitting can be neglected for high field strength \cite{delone_krainovII}. In this high-frequency limit the shift of the ionization potential $I_p$ is equal to the shift of highly excited states. In contrast, the ground state satisfies $E_g \gg \omega$ and the corresponding shift is given by the quasi-static Stark shift and is much smaller than the ponderomotive shift \cite{delone_krainovII}.

In general, the modification of the electronic structure of atoms and molecules due to the interaction with a strong laser field is ubiquitous, and manifests in several aspects of laser driven physics, for example laser filamentation \cite{richter2013role, matthews2018amplification}, photo-electron spectroscopy \cite{morales2011imaging} or atomic stabilization \cite{fedorov1990coherence, dubrovskii1991resonant, ivanov1994suppression}.
In particular, the effects associated with the ponderomotive shift of the ionization potential and the excited states lead to channel closing \cite{schyja1998channel} and the observation of Freeman resonances \cite{freeman}, respectively. In this work, the appearance of Freeman resonances is of particular interest. They occur when an excited state is shifted in such a way that ionization from the ground state is now resonant with the Stark shifted excited state. Thus additional peaks in the photo-electron energy distribution can be observed \cite{freeman, perry1989resonantly, wiehle2003dynamics}. 
This resonant ionization is strongly suppressed in circular polarization \cite{freeman_circular}. 
A related perspective on laser dressed resonant ionization pathways is given by the concept of frustrated tunnel ionization (FTI) \cite{eichmann_FTI, zimmermann2015strong, zimmermann2016atomic, popruzhenko2017quantum}. FTI describes the trapping of electrons in excited states during optical tunneling, with classical (Kepler) orbit arguments. For certain initial velocity conditions, these electrons are trapped in Rydberg orbits around the core, which, in turn, survive the interaction with the strong laser field \cite{eichmann2013observing}. Frustrated tunneling only occurs in linear polarization, since circular polarized fields do not lead to such closed orbits \cite{eilzer2014steering}.

In recent years much attention has been given to two-color circular polarized laser fields in strong field physics \cite{milovsevic2000generation, milosevic_spin, ayuso2017attosecond, jimenez2018control, eicke2019attoclock, abu2019pulse}. These so-called bi-circular laser fields consist of two co-planar circular polarized laser fields which either co- or counter-rotate with respect to each other, and are usually set with an integer $s$ frequency ratio between both fields, i.e. $s \omega_1 = \omega_2$. The case of counter-rotating fields is of particular interest since the total electric field vector gives rise to a Lissajous figure which changes in magnitude and direction and therefore combines properties of both linear and circular polarized fields. These properties are then imparted on the trajectory of the electrons in the continuum after ionization. Thus processes only typical of linear polarization, such as high harmonic generation (HHG), (in)-elastic scattering at the Coulomb potential \cite{ivanov_krausz_review}, or intra-cycle (sub-cycle) interference \cite{arbo2010intracycle} combined with processes known for circular polarization, such as the generation of spin-polarized electrons \cite{barth2013spin,hartung2016electron,eckart2018ultrafast}, have been observed in bi-circular laser fields \cite{mancuso2016controlling, mauger2016circularly, ayuso2017attosecond, jimenez2018control,eckart_subcycle}. Other recent works have also focused on how the 3-fold symmetry that the bi-circular field imposes, is broken in the presence of the Coulomb potential \cite{gazibegovic2018helicity}.

In this work we report on the appearance of AC-Stark shifted resonances in bi-circular fields, and how it is possible to obtain information about the electron dynamics in Rydberg states prior to ionization. This is described in terms of interference between a direct and a resonant ionization channel, and is manifested in new angular structures in the low energy region of the photo-electron spectra. We further report on how the Rydberg state population, responsible for the low energy structures, change with varying pulse duration and intensity.

\section{\label{sec:methods}Methods}

The time-dependent Schrödinger equation (TDSE) was solved numerically for the hydrogen atom and a short range Yukawa potential using the SCID-TDSE code \cite{patchkovskii_TDSE}, that utilizes the typical partial wave expansion of the wavefunction, exploiting the spherical symmetry of the problem. It propagates the wave-function using a rotating reference frame, which allows for calculations in arbitrary fields, in a tractable amount of time \cite{patchkovskii_TDSE}. The Hamiltonian used in these calculations is within dipole approximation and in the velocity gauge. The photo-electron momentum distributions have been calculated using the iSURF method \cite{morales_tsurf}, which is an extension of surface flux methods \cite{ermolaev1999integral, ermolaev2000integral, scrinzi_tsurf}, that allows for a projection to infinite time. In the case of the Hydrogen atom calculations, the asymptotic functions chosen to project the wavefunction were Coulomb waves (iSURFC), while for the case of the Yukawa potential, it was performed using plane waves (iSURFV).
For the simulation box used in the calculations, we employ a non-uniform grid of 790$a_0$ (with the Bohr radius $a_0$), with the first 10 points having a linear spacing of 0.3636$a_0$, followed by 25 points with a logarithmic scaling, and a parameter of 1.1, starting at 0.4$a_0$, finishing with a 1965 point uniform grid with spacing of 0.4$a_0$. 
We place a transmission-free Complex Absorbing Potential \cite{manolopoulos2002} at $R \sim$ 756$a_0$. The temporal discretization has a time-step of 0.0025 atomic units. The calculations include angular momentum up to $l = 60$ and $m = \pm 60$, and convergence was checked with respect to all parameters.

\begin{figure}
    \centering
    {\includegraphics[width=7.5cm]{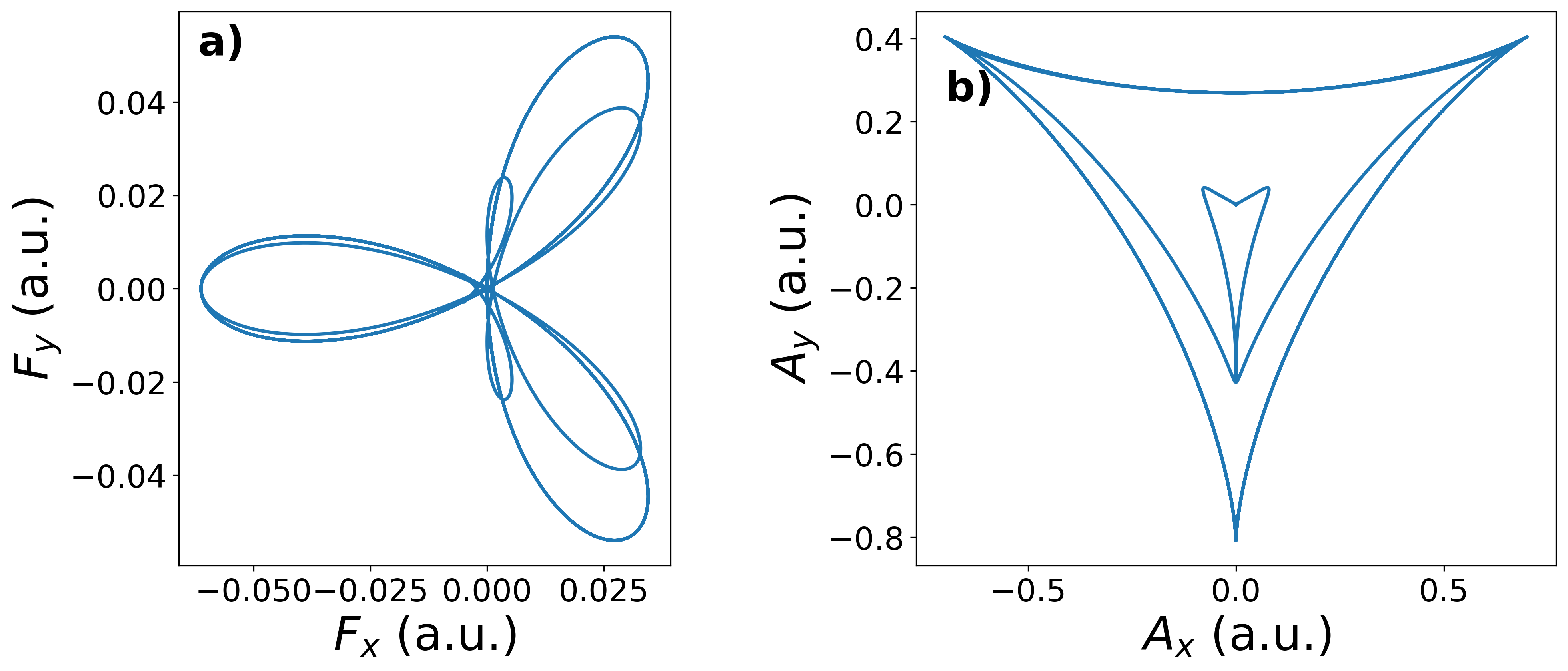}}
      \caption{Electric field (a) and vector potential (b) from Eq. (\ref{eq:field}) for a field strength of $F_0=0.03065$ a.u. and a six cycle pulse ($t_{on}=t_{off}=1$ cycle, $t_{flat}=4$ cycles).}
      \label{fig:fields}
\end{figure}

The system was exposed to an intense bi-circular laser field (see Fig.\ref{fig:fields}) with a fundamental field of frequency $\omega = 0.0569$ atomic units and its counter-rotating second harmonic. The vector potential of the field $\vb{A}(t)$ in the xy-polarization plane is given by (atomic units are used throughout the paper)  
\begin{equation}
\label{eq:field}
\begin{aligned}
A_x(t) &= A_0[-\sin(\omega t) - \frac{1}{2} \sin(2 \omega t)] f(t), \\
A_y(t) &= A_0[\cos(\omega t) - \frac{1}{2}\cos(2 \omega t)] f(t),\\
\end{aligned}
\end{equation}
where $A_0 = F_0 /\omega$ is the vector potential amplitude, which is expressed in terms of the electric field strength $F_0$ and $f(t)$ is the envelope. In all calculations the field strength $F_0$ was equal for both fields and varied in a range from $0.018 - 0.035$ atomic units, corresponding to peak intensities of $4.5 \times 10^{13} - 1.7 \times 10^{14}$ W/cm$^2$. In order to avoid envelope effects on the 3-fold symmetry of the photo-electron spectra, we use a flat-top envelope 

\begin{equation}
\label{eq:envelope}
\begin{aligned}
f(t) = \begin{cases}
\sin^2\left(\frac{\omega t}{4 N_{on}}\right), & 0 < t \le t_{on} \\
1, & t_{on} <  t < t_{on} + t_{flat} \\
\sin^2\left(\frac{\omega (t - t_{flat})}{4 N_{off}}\right), & t_{off} + t_{flat} \le t \le \tau \\
0, &otherwise
\end{cases}
\end{aligned}
\end{equation}
where we demand $N_{on} = N_{off} \equiv N$ and $t_{on/off} = 2\pi N /\omega$ indicates the number of cycles and the duration of the envelope turn-on and turn-off. For $N_{flat}$ cycles in the flat-top region with duration $t_{flat} = 2\pi N_{flat} /\omega$, the total pulse duration is $\tau = 2\pi (2N + N_{flat})/\omega$. For the turn-on and turn-off one cycle was used. The flat-top region was varied from 1 to 40 cycles ($t_{flat} \approx 2.7 - 108$ fs). The electric field $\vb{F}(t)$ and the corresponding vector potential $\vb{A}(t)$ from Eq.(\ref{eq:field}), are shown in Fig.\ref{fig:fields}. 

The short range Yukawa potential was investigated to understand the effect of the Coulomb potential, and is given by 
\begin{equation}
\label{eq:yukawa}
U(r) = -\frac{Z}{r} e^{-r/a},
\end{equation}
with $Z$ and $a$ being the effective charge and radius of the potential, respectively. The Yukawa potential has the advantage of supporting only one bound state and with the particular choice of $Z = 1.94$ and $a=1.0$ this ground state has the same ionization potential like hydrogen, namely $I_p \approx 0.5$ a.u., and thus resonance effects in the Yukawa system are eliminated.

\section{\label{sec:results}Results and discussion}

\begin{figure}
	\includegraphics[width=7.0cm]{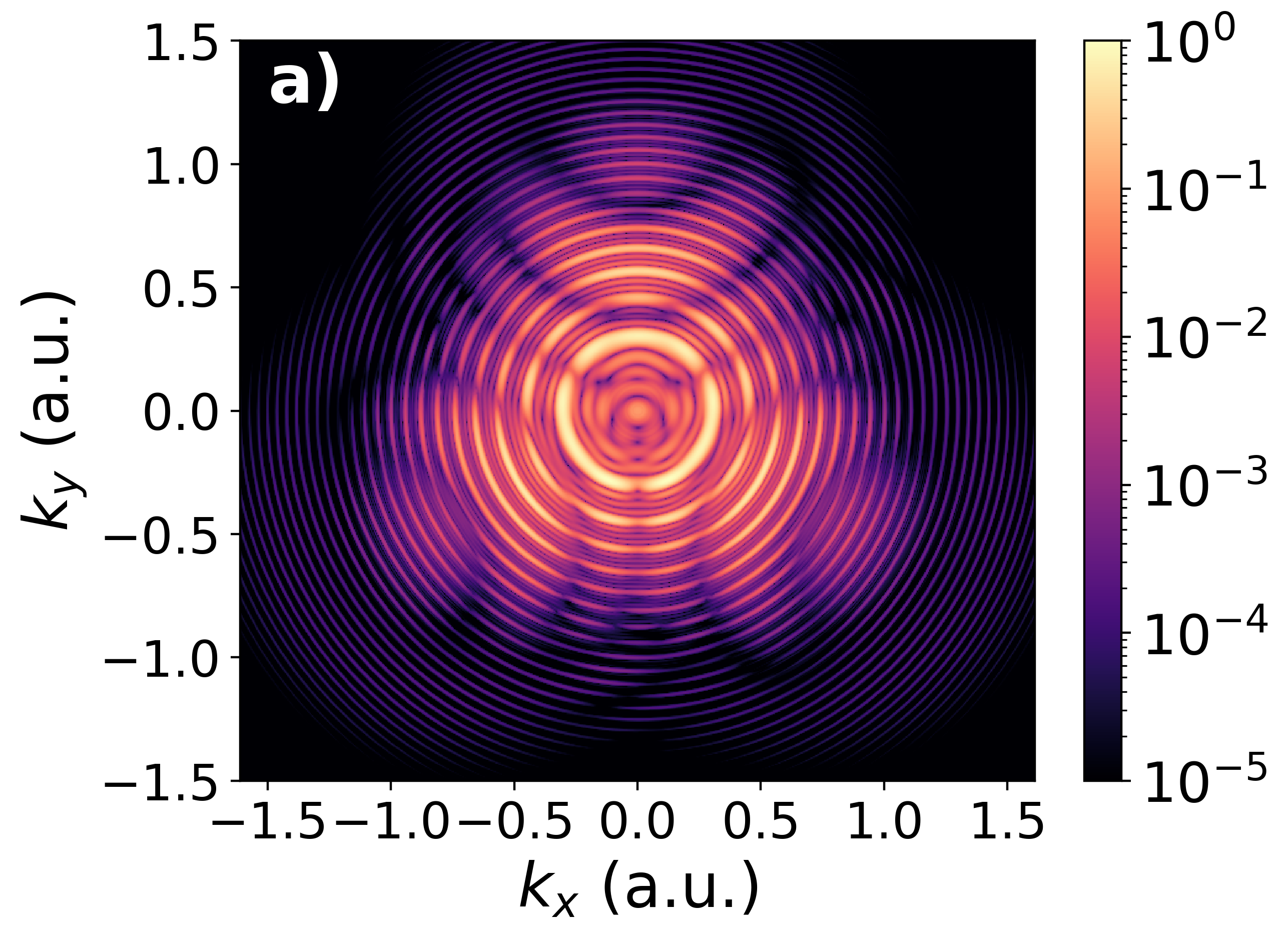}
	\includegraphics[width=7.0cm]{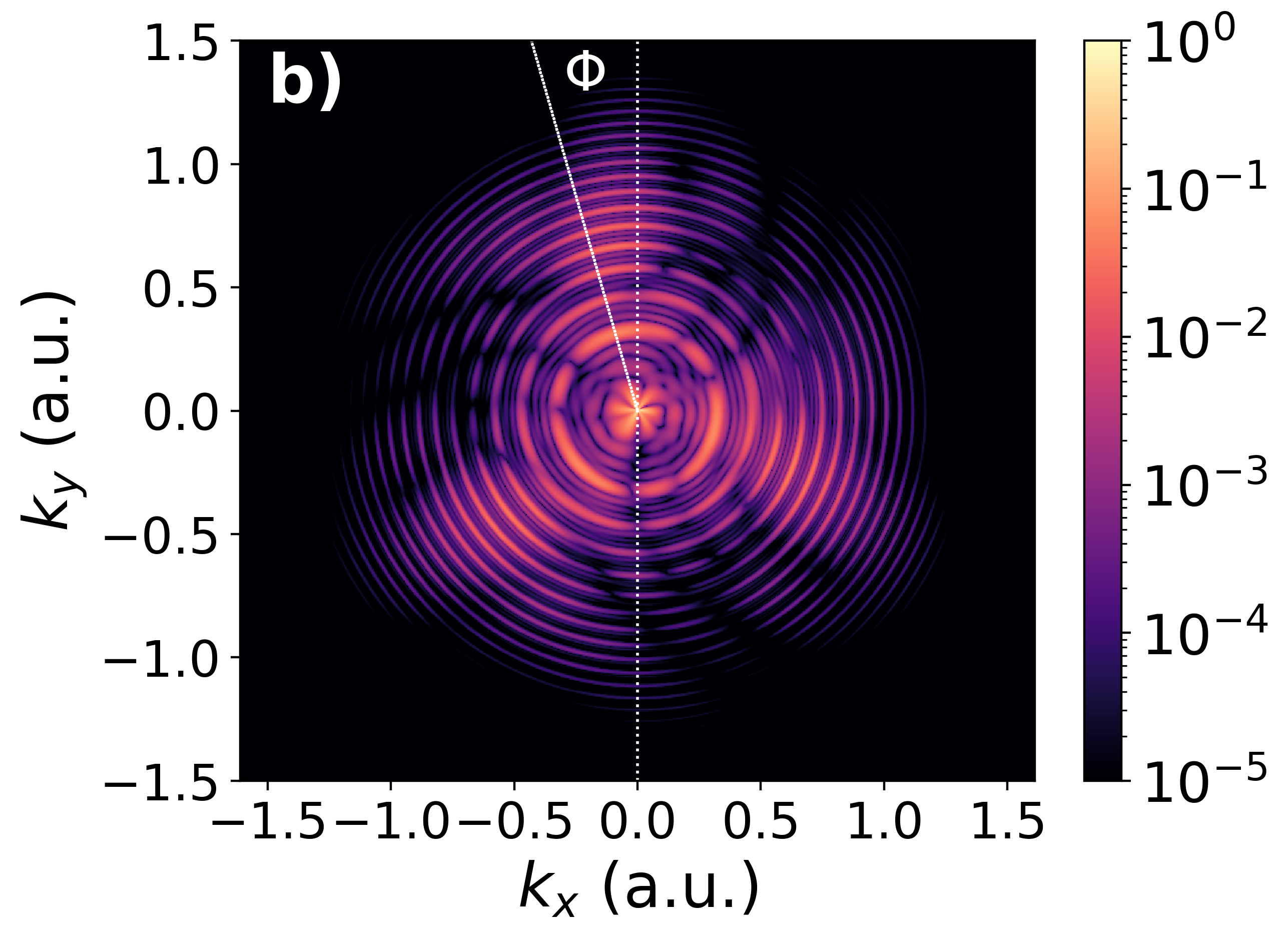}	
	\caption{Photo-electron momentum distribution after strong field ionization in a bi-circular laser field normalized with respect to the highest intensity on logarithmic scale: from the short range Yukawa potential (a) and hydrogen (b). A 1-4-1 cycle pulse ($t_{on}=t_{off} = 1$ cycle, $t_{flat} = 4$ cycle, i.e. $\tau \approx 16$ fs) with the field strength of $F_0 = 0.03065$ atomic units was used. The offset angle $\Phi$ of the predominant propagation direction in the hydrogen atom with respect to the Yukawa potential is indicated in (b).}
	\label{fig:spectra1}	
\end{figure}

The calculated photo-electron momentum distribution in the polarization plane of the bi-circular laser field for the Yukawa and the Coulomb potential for a field strength of $F = 0.03065$ a.u. and a six cycle pulse (1 cycle turn on, 4 cycles flat, 1 cycle turn off, $\tau \approx 16$ fs) are shown in Fig.\ref{fig:spectra1}. 

The angular distribution of the spectra for the Yukawa case (\ref{fig:spectra1}a), clearly follows the overall 3-fold symmetry in the ATI rings (symmetric under rotations of $120$ degree) imposed by the vector potential, except at small momenta. Due to the absence of the long range Coulomb potential the predominant propagation direction of the electrons is at the angles of $90, 210, 330$ degree , i.e. in the negative direction of the vector potential (compare with Fig.\ref{fig:fields}), as the electron momentum after ionization is mainly dictated by the vector potential $\vb k = - \vb A(t_0) $ at the instant of ionization $t_0$ \cite{ivanov_krausz_review}. This instant of ionization is given by the time where the field strength reaches its maximum, marking the instant of highest ionization probability due to the exponential dependence of the tunnel probability on the barrier width (note that the electric field and the vector potential are orthogonal at all times).

Replacing the Yukawa with the Coulomb potential reveals new features in the spectra. The predominant direction in the photo-electron momentum distribution is now shifted by an offset angle $\Phi(|\textbf{k}|)$ (Fig.\ref{fig:spectra1}b), that depends on the momentum of the photo-electron $|\textbf{k}|$. 
This is a consequence of the interaction of the ionized electron with the long range Coulomb potential during the propagation of the electron in the continuum driven by the electric field, in the same way direct electrons drift in the case of circularly polarized light \cite{attoclock_morales}. Direct electrons with different momentum (i.e. different ATI peaks) drift towards the detector with different velocities, and therefore, feel the Coulomb potential differently: slower electrons will be deflected more than the faster ones. 
This angle of deflection also depends on the intensity of the electric field. This dependence on the photo-electron momentum $|\textbf{k}|$ has previously been observed in single-color elliptical polarized \cite{offset_elliptical} and bi-circular fields \cite{madsen_TDSE_bicircular}. In \cite{madsen_TDSE_bicircular} the authors show that for the first few ionization peaks, the offset angle does not obey the behavior known from the direct electrons in single-color circular polarized fields, i.e. that the offset angle decreases with increasing intensity. However, in our results, we observe that this offset angle dependence, considering the influence of direct electrons alone, does indeed follow the same trend as that of the single-color circular case.

The other remarkable feature that arises in the presence of the Coulomb potential is the flower-like pattern that appears in the low energy region of the spectra. We will show that this six-fold structure appears due to the presence of additional bound states and the corresponding population dynamics prior to ionization. We will also show how these structures appear from the interference between direct and resonant (via AC-Stark shifted states) ionization channels.

\begin{figure}
	\centering
	\includegraphics[width=7.0cm]{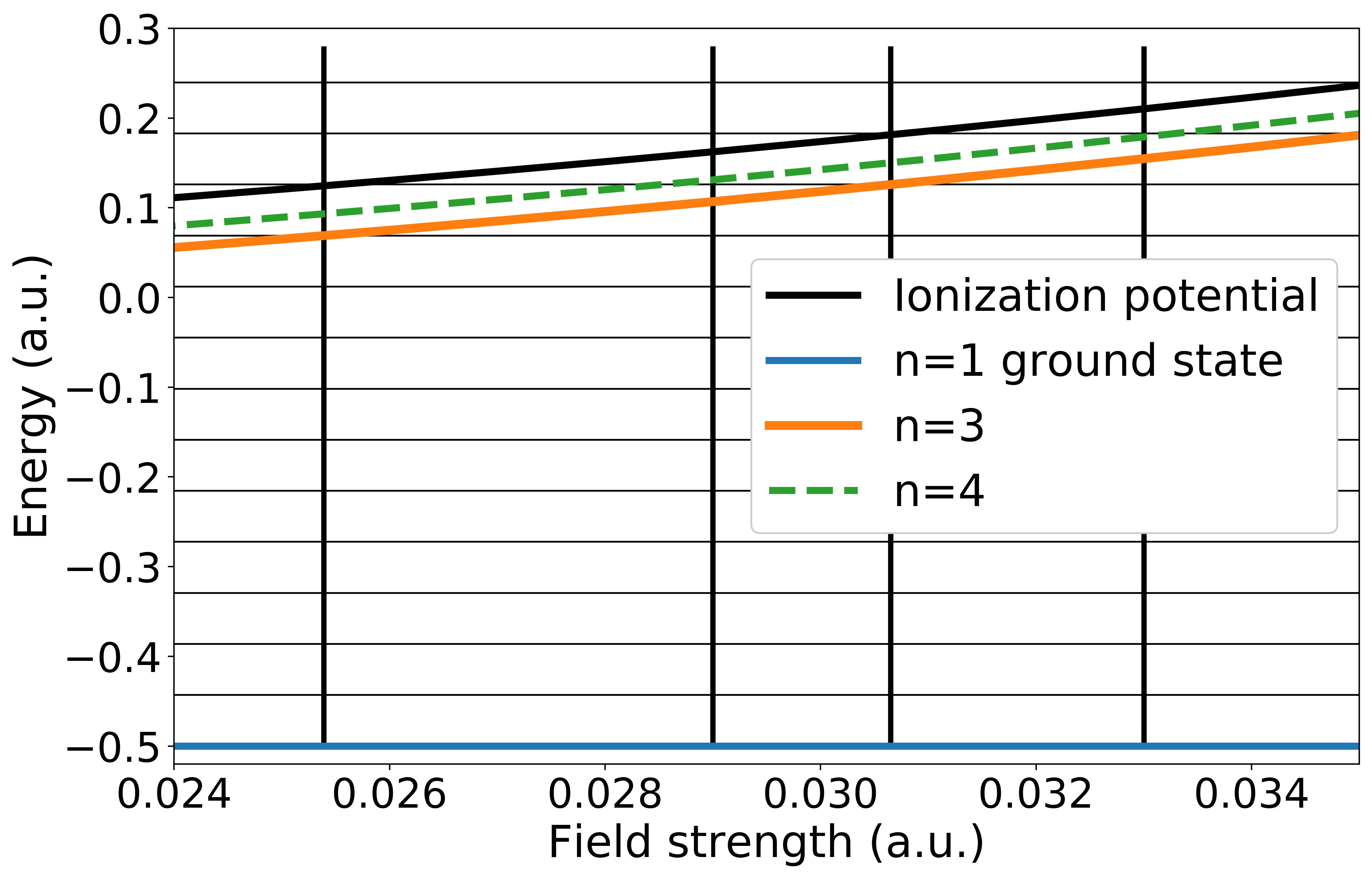}
	\caption{Energy diagram of the $U_p$-shifted $n=\{3,4\}$ excited hydrogen eigenstates from Eq.(\ref{eq:AC_stark}). The horizontal lines indicate the ground state energy dressed by up to 13 $\omega$ photon energies of $E_{ph} = 0.0569$ a.u. The spectra from Fig.\ref{fig:spectra2} (a)-(d) are obtained at a field strength indicated by vertical lines.}
	\label{fig:UP_shift}
\end{figure}

In order to pursue the origin of this structure, let us first consider the energy levels of the total system, namely the hydrogen atom plus the bi-circular laser field. The previously introduced condition for the dynamical Stark shift in the high frequency limit of excited states, i.e. $\omega \gg n^{-3}$ for a Rydberg state with principal quantum number $n$, is already applicable for states with $n \ge 3$ for the typical fields used in these calculations. Therefore we consider that the hydrogen eigenstates are shifted by the ponderomotive energy $U_p$. Since the hydrogen ground state satisfies $|E_g| \gg \omega$, the ponderomotive shift can be neglected \cite{delone_krainovII}. The corresponding field strength dependent bound state energies $E_{n}(F_0)$ are shown in Fig.\ref{fig:UP_shift} and given by 

\begin{equation}
\label{eq:AC_stark}
\begin{aligned}
E_{n} (F_0) &= E_n^{0} + U_p(F_0), \\ 
U_p (F_0)  & = \frac{1}{T} \int_0^T \frac{1}{2} [\dot x(t)^2 + \dot y(t)^2] dt =  \frac{F_0^2}{2\omega^2}+ \frac{F_0^2}{8 \omega^2},
\end{aligned}
\end{equation}  
with the unperturbed hydrogen eigenenergies $E_n^{0} = - 1/2n^2$ with principal quantum number $n \ge 3$ and the field strength dependent ponderomotive energy $U_p(F_0)$ for the field defined by Eq.(\ref{eq:field}) and fundamental frequency $\omega$. The velocities $\dot x(t)=A_0[-\sin(\omega t)- \frac{1}{2}\sin(2\omega t)]$ and $\dot y(t) = A_0[\cos(\omega t)- \frac{1}{2}\cos(2\omega t)]$ in the polarization plane are obtained from classical equations of motion for an electron in an oscillating laser field with the vector potential from Eq.(\ref{eq:field}).

The ground state, neglecting it's AC-Stark shift, dressed by up to 13 photons of the fundamental field ($\omega = 0.0569$ a.u.) is shown in Fig.\ref{fig:UP_shift} with horizontal lines, where the horizontal axis denotes the field strength, and the vertical axis the energy of the Stark shifted states. For the field strength of Fig.\ref{fig:spectra1} ($F_0=0.03065$ a.u.), the $n=3$ hydrogen eigenstate is AC-Stark shifted into resonance with the ground state, dressed with 11 photons of the fundamental field. From this $n=3$ state the absorption of one photon from the fundamental field $\omega$ is sufficient to ionize the excited atom. Thus single photon ionization pathways from the excited state and multi-photon ionization pathways from the ground state will interfere at the first ATI peak of the spectra. 

\begin{figure}
	\includegraphics[width=4.2cm]{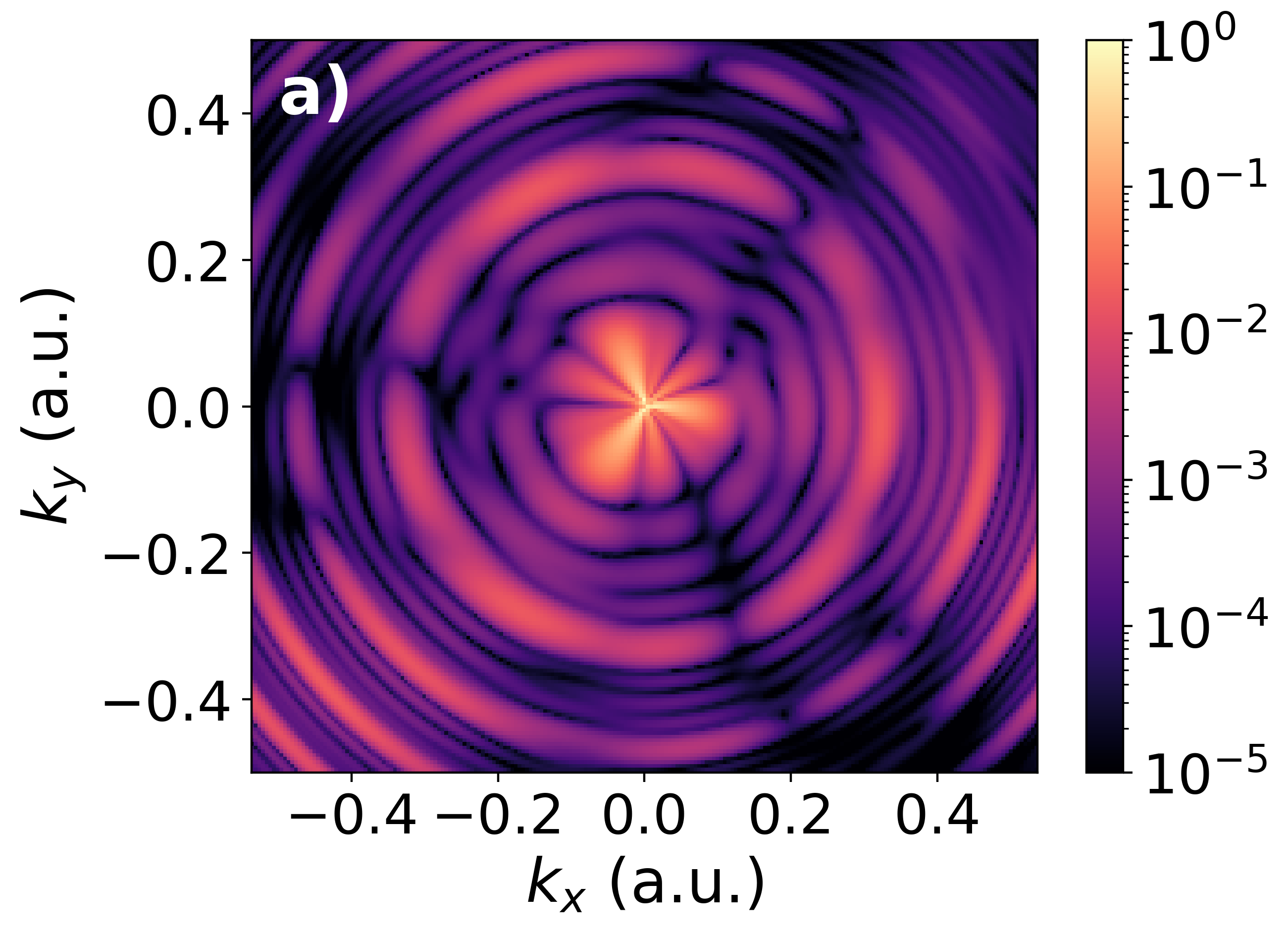}
	\includegraphics[width=4.2cm]{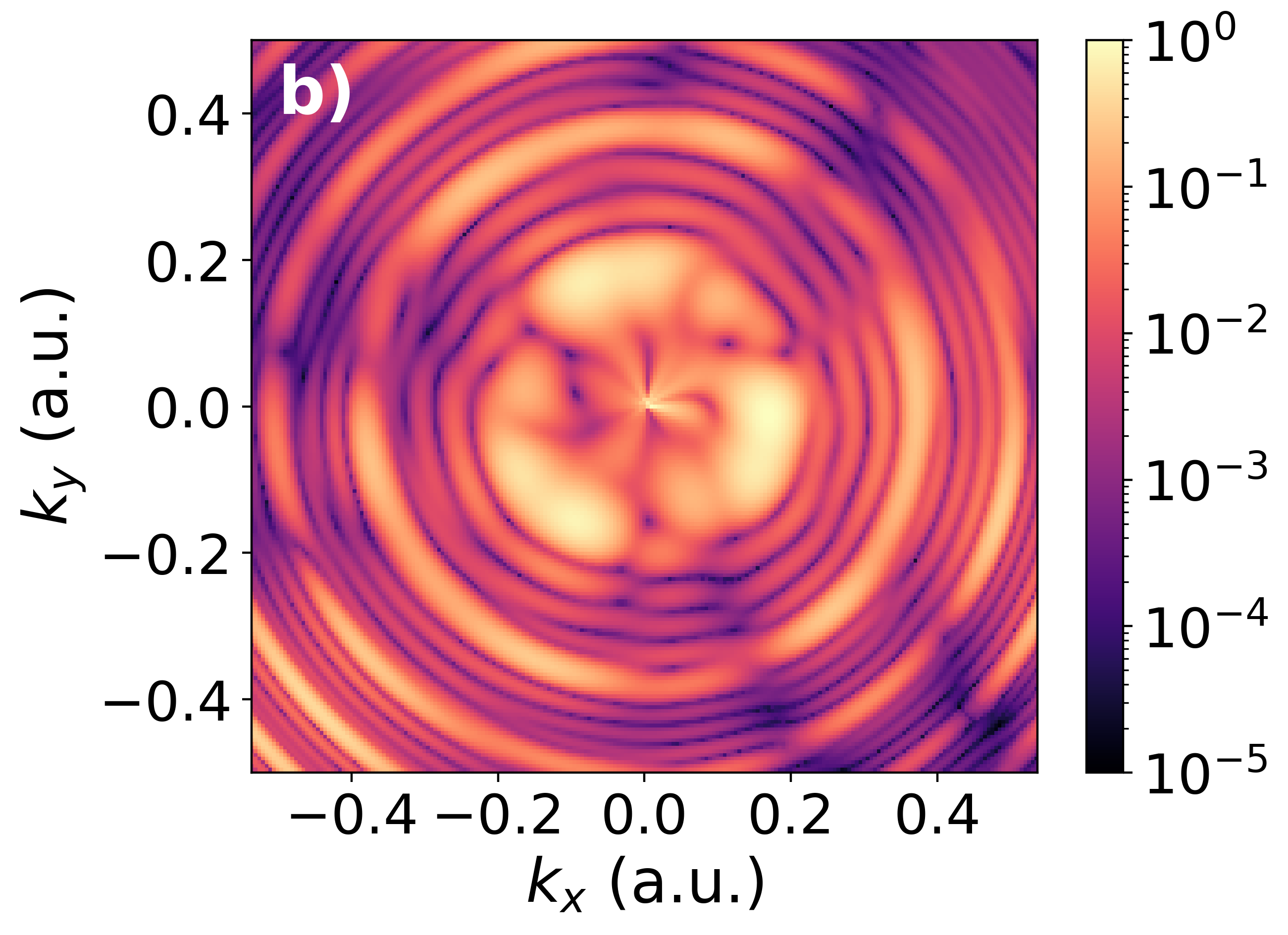}
	\includegraphics[width=4.2cm]{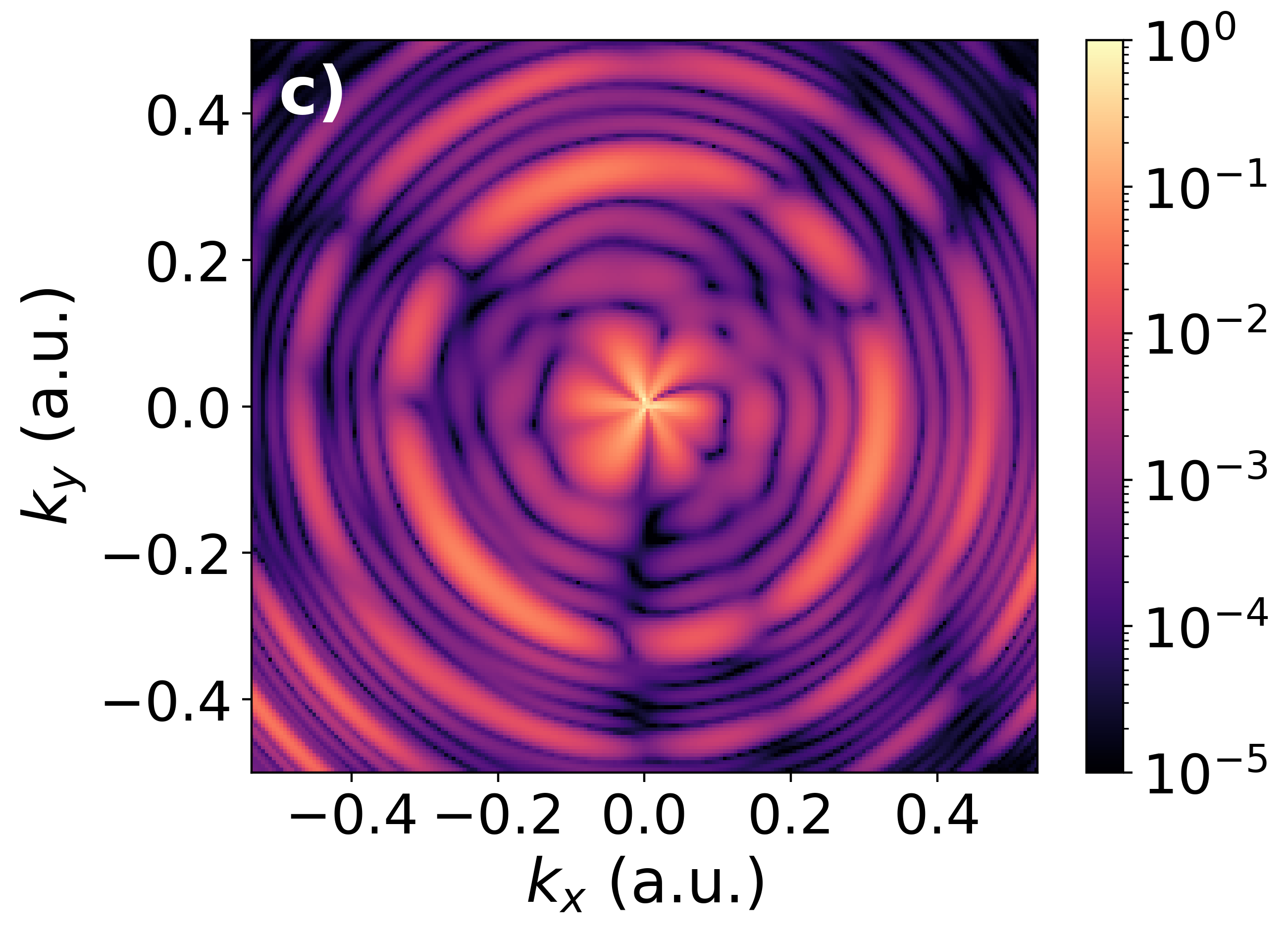}	
	\includegraphics[width=4.2cm]{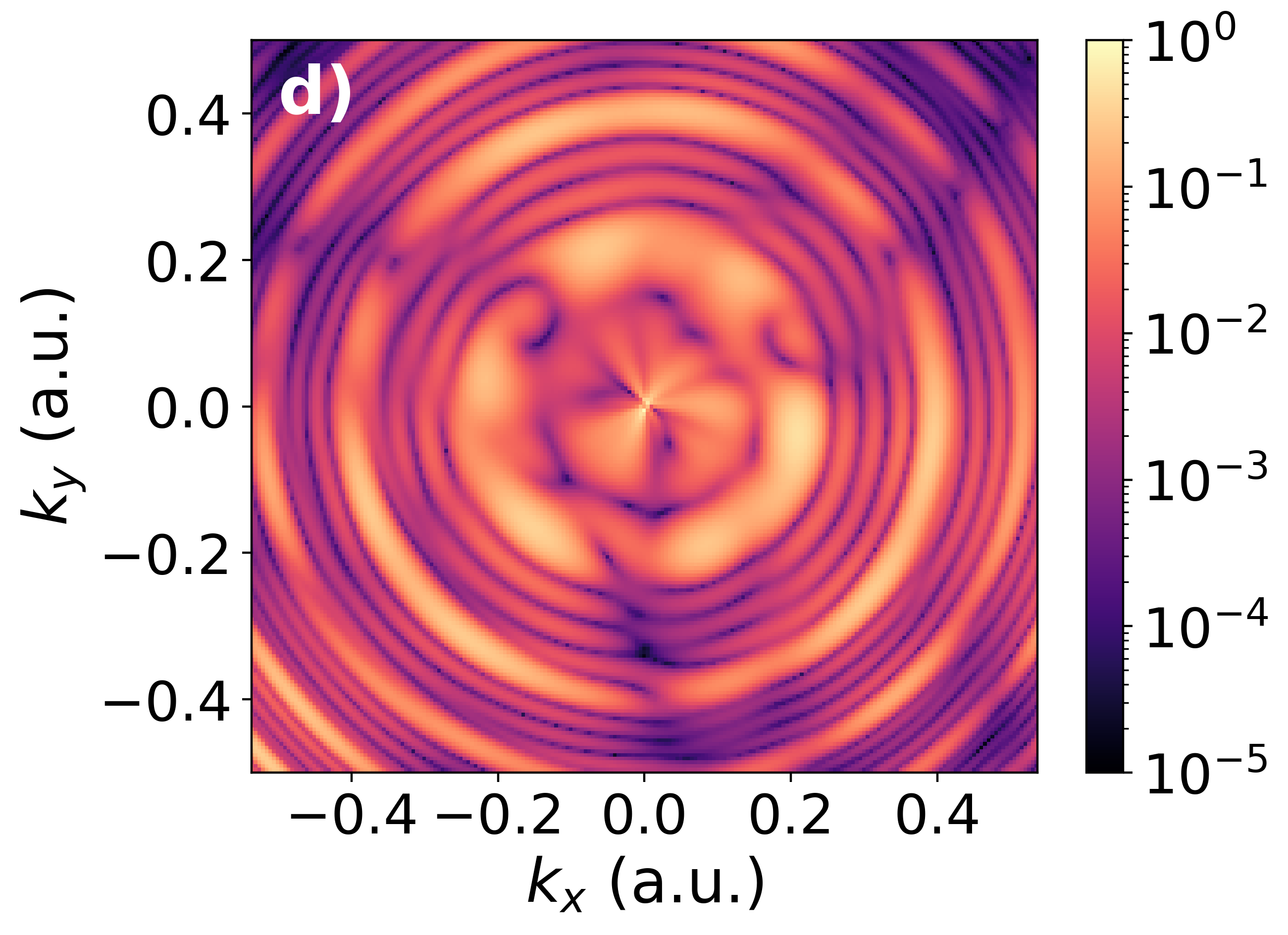}
	\caption{Low energy region of the normalized phot-electron momentum distribution after strong field ionization in a bi-circular laser field for the hydrogen atom, for different field strength $F_0 = 0.0254$ (a), $0.029$ (b), $0.03065$ (c) and $0.033$ (d) atomic units with a six cycle pulse (1-4-1 pulse as described in Fig.\ref{fig:spectra1}, $\tau \approx 16$ fs).}
	\label{fig:spectra2}
\end{figure}

In order to confirm this ionization pathway interference mechanism as the origin of the observed low energy structure we show the low energy region of the photo-electron momentum distribution for four different field strength ($F_0 = 0.02539,0.029,0.03065, 0.033$ a.u.) in Fig.\ref{fig:spectra2}. For the field strength $F_0 = 0.0254$ a.u. and $F_0=0.03065$ a.u. (Fig.\ref{fig:spectra2}, (a) $\&$ (c)) the $n=3$ AC-Stark shifted excited state is resonant with the ground state dressed by 10 and 11 photons, respectively. The 6-fold structure in the photo-electron spectra at low energies shows similar symmetric pattern for both cases. On the other hand, for the field strengths of (b) and (d) in Fig.\ref{fig:spectra2}, the distinct structure disappears, as the $n=3$ AC-Stark shifted eigenstate is no longer in resonance with the dressed ground state (see Fig.\ref{fig:UP_shift}). It is apparent in the spectra, that the effect of the resonant pathway is switched on and off, and the 6-fold structure at low energy almost disappears. As the intensity changes, the $n=3$ states moves in and out of resonance, as it is expected from a Freeman resonance. However, it should be noted that the observed features in the low energy region are not dictated by the three-fold symmetry of the laser field as expected for tunnel-ionization processes. This suggest that the underlying mechanism of this low energy structure is indeed dominated by multi-photon processes. For the off-resonant cases in Fig.\ref{fig:spectra2} (b) and (d) one can discern six-fold symmetric structures ($|k| \sim 0.2$ a.u.). But, this feature is not fully invariant under rotations of $60$ degree, but, it is however invariant under rotations of $120$ degree. Thus we are in a competing regime between tunnel-ionization and multi-photon ionization, which is supported by a Keldysh parameter of $\gamma = \sqrt{I_p / 2U_p} = 1.1 - 2.0$, for the field strength under consideration. In the resonant cases, however, the multi-photon excitation and ionization mechanism dominates the spectra for low energies.

\begin{figure}
	\includegraphics[width=4.2cm]{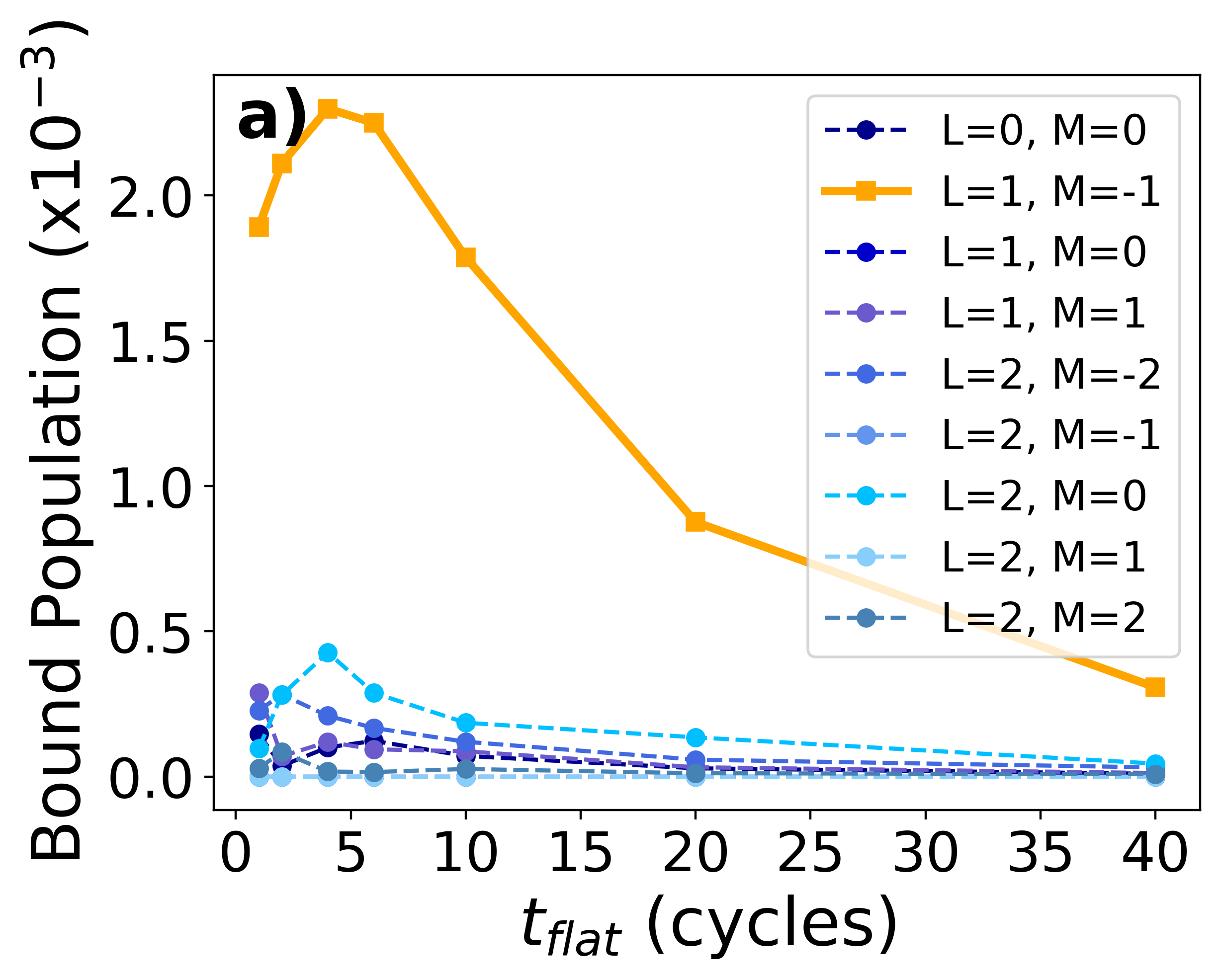}
	\includegraphics[width=4.2cm]{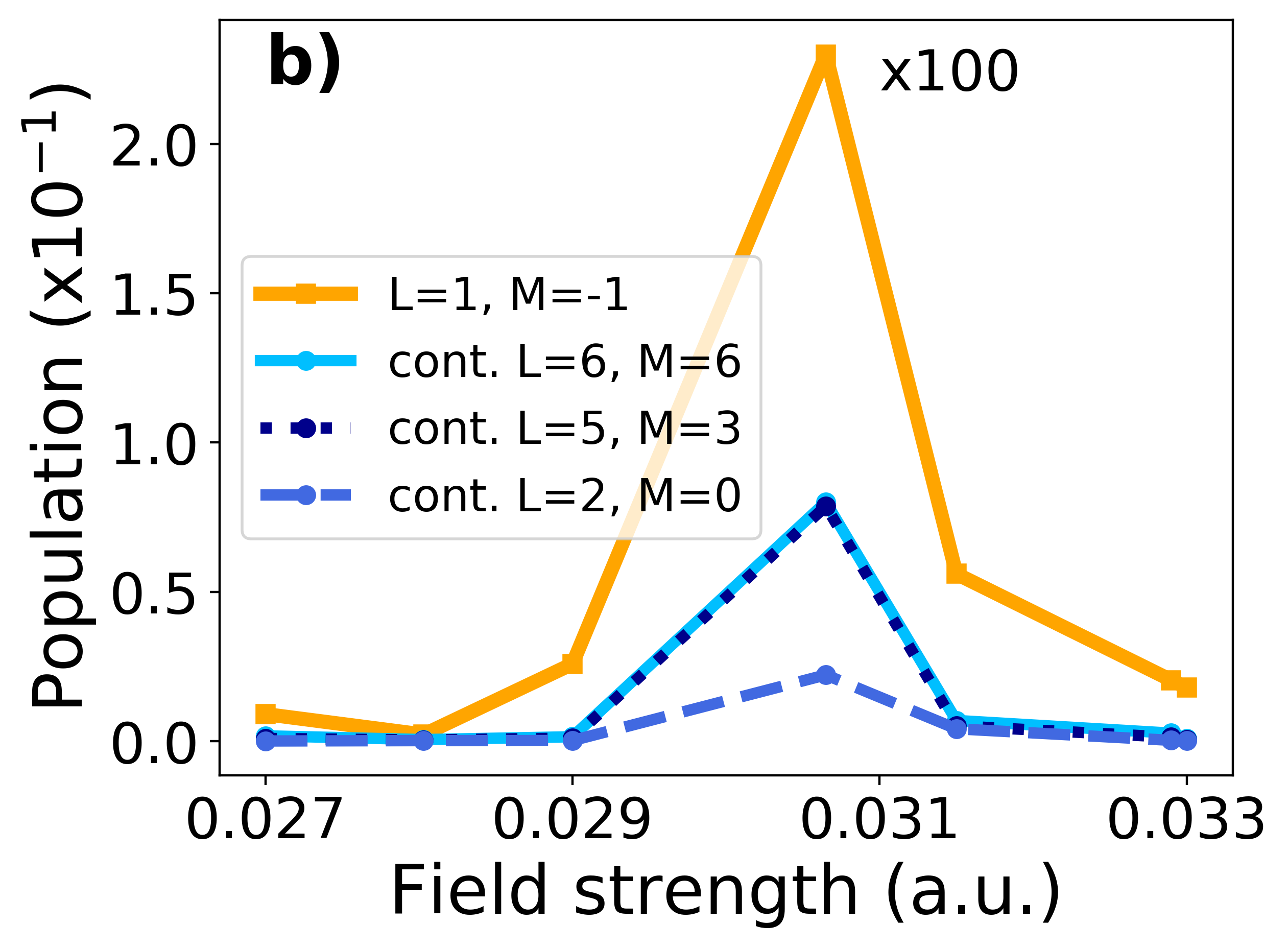}
	\caption{Absolute population of bound and continuum states obtained from the solution of the TDSE. a) The $n=3$ bound resonant states are resolved in their angular momentum $l$ and $m$ for different pulse duration ($1 \le t_{flat} \le 40$) and fixed field strength of $F_0 = 0.03065$ atomic units. The most populated bound state ($l=1,m=-1$, solid) is resonantly excited via absorption of 11 photons ($3 \times  \omega \, + \, 4 \times 2 \omega$ photons, according to dipole selection rules). b) Population of the most populated bound state $l=1,m=-1$ of the $n=3$ manifold (squares, solid) for different field strength with a $t_{flat} = 4$ cycle pulse, with the characteristic increase of the population at the Freeman resonance, peaking at the resonance field strength of $F_0 = 0.03065$ atomic units. The two most populated continuum states ($l=6,m=6$, circles, solid and $l=5,m=3$, circles, dotted) and the continuum state for the resonant ionization path ($l=2,m=0$, dashed) peak at the resonance field strength. This resonant behavior originates from the bound resonance ($l=1, m=-1$) and continuum resonance (at this particular $|k|$) for the corresponding resonant and direct ionization pathways, respectively.}
	\label{fig:population1}
\end{figure}

But what is the origin of this 6-fold structure? It can be understood by carefully analyzing the partial wave population of the continuum waves for the particular final momentum region where this structure shows up and how it correlates with the bound state population.  Therefore we further analyze the population of the bound and continuum states after the end of the pulse, for the particular case of the $n=3$, 11-photon resonance ($F_0 = 0.03065$ a.u., Fig.\ref{fig:spectra2}(c)). The continuum momentum at the detector where the low energy structure appears, ($|k| < 0.05$ a.u.) is accessible via the absorption of 12 photons with energy $E_{ph} = 0.0569$ a.u. from the initially populated ground state (direct pathway), or via the absorption of one $\omega$ photon from the $n=3$ excited state (resonant pathway). Note that absorption of two $\omega$ photons from the $n=3$ bound state (or one $2\omega$ photon) will be out of the range where the structure dominates. Both ionization paths lead to the same final momentum and interfere at the detector. For a deeper understanding of the underlying population mechanism, we analyze the bound and continuum populations at the end of the pulse, shown in Fig.\ref{fig:population1}, for different pulse duration and varying field strength, from the solution of the TDSE (resolved in angular momentum $l$ and $m$). 

The dominant contribution to the $n=3$ bound manifold, Fig.\ref{fig:population1}(a), is given by the state with angular momentum $l=1,m=-1$ (solid). It can be populated by the absorption of $3$ and $4$ photons from the fundamental field $\omega$ and its second harmonic $2\omega$, respectively, consistent with dipole selection rules. The decrease in the total $n=3$ yield for long pulses can be understood in terms of FTI \cite{eilzer2014steering,zimmermann2017unified,ortmann2018controlling}, and will become clear when investigating the quantum number $n$ distribution for different pulse duration . 
Out of the continuum states (Fig.\ref{fig:population1}(b), circles), represented by the angular momentum $l$ and $m$ and their momentum $|k|$, we  focus on the states within the low momentum region $|k| \le 0.2$ a.u., and find that the two most populated continuum states are $l=6,m=6$ (circles, solid) and $l=5,m=3$ (circles, dotted). These states can be reached from the ground state by absorption of $8 \times \omega \, + 2 \times 2\omega$ and $6 \times \omega \, + 3 \times 2\omega$ photons respectively, from the two fields according to dipole selection rules.  Additionally we will focus on the contribution of the $l=2,m=0$ continuum state (circles, dashed). First, it is the particular continuum state with the highest relative population which can be reached by any of the $n=3$ bound states by the absorption of a single $\omega$ photon in order to result in electrons which remain within the region of interest ($|k| < 0.05$ a.u.). Furthermore, it is this particular continuum state which is reached from the highest populated resonant bound state ($l=1,m=-1$) of the $n=3$ Rydberg state manifold by single photon absorption of the fundamental field $\omega$. 

In Fig.\ref{fig:population1}(b) the population of the most important populated bound and continuum states are shown for various field strength around the 11-photon resonance. The population of the bound state with $l=1,m=-1$ (squares, solid) clearly peaks at the resonant field strength indicating the resonant population enhancement. The population of continuum states (circles) is shown for the two dominant states, i.e. $l=6,m=6$ (solid) and $l=5,m=3$ (dotted), and for the continuum state $l=2,m=0$ (dashed). The peak of the $l=2,m=0$ continuum state participating in the resonant ionization pathway can be understood by the effect of enhanced bound population and subsequent ionization. On the other hand, the peak corresponding to the two states of the direct ionization channels seems surprising since they are not directly correlated to the resonant excitation of the $n=3$ bound state. Because we are focusing on the population of these continuum channels at a given momentum ($|k| \le 0.03$ a.u.), this population is also affected by the AC-Stark shift of the ionization threshold, giving rise to their population enhancement for this particular field strength.

We now provide a simple model in order to understand the shape of the AC-Stark shifted resonance manifestation in the photo-electron momentum distribution. In order to associate this resonant population enhancement with the observed spectral feature, we will consider a simple interference model for the different ionization pathways. In particular taking into account the most populated continuum states for the direct and resonant ionization channels, represented by the spherical harmonics $Y_{6,6}$ and $Y_{2,0}$, respectively, we can reproduce the 6-lobe structure from the the low energy region. Since the $Y_{6,6}$ continuum state can not be reached via single photon absorption from any $n=3$ bound state by dipole selection rules, this continuum state is associated with direct ionization. In contrast, the $Y_{2,0}$ continuum state can be reached via single photon ionization by the fundamental field $\omega$ from the most populated bound state $Y_{1,-1}$ of the $n=3$ manifold (see Fig. \ref{fig:population1}). This is considered as the resonant ionization path. Due to the indistinguishability of the two pathways, the direct and the resonant ionization path will interfere at the detector. Since the state of the indirect path interacts with the excited bound state, it can potentially acquire a relative phase $\chi$ with respect to the state of the direct ionization channel, such that the total wavefunction in the continuum is given by
 
\begin{equation}
\psi(\theta, \phi; \chi) = Y_{6,6}(\theta, \phi) + e^{i\chi} Y_{2,0}(\theta, \phi).
\label{eq:sph_harm_sum}
\end{equation}  
 
\begin{figure}
	\centering
	\includegraphics[width=8.0cm]{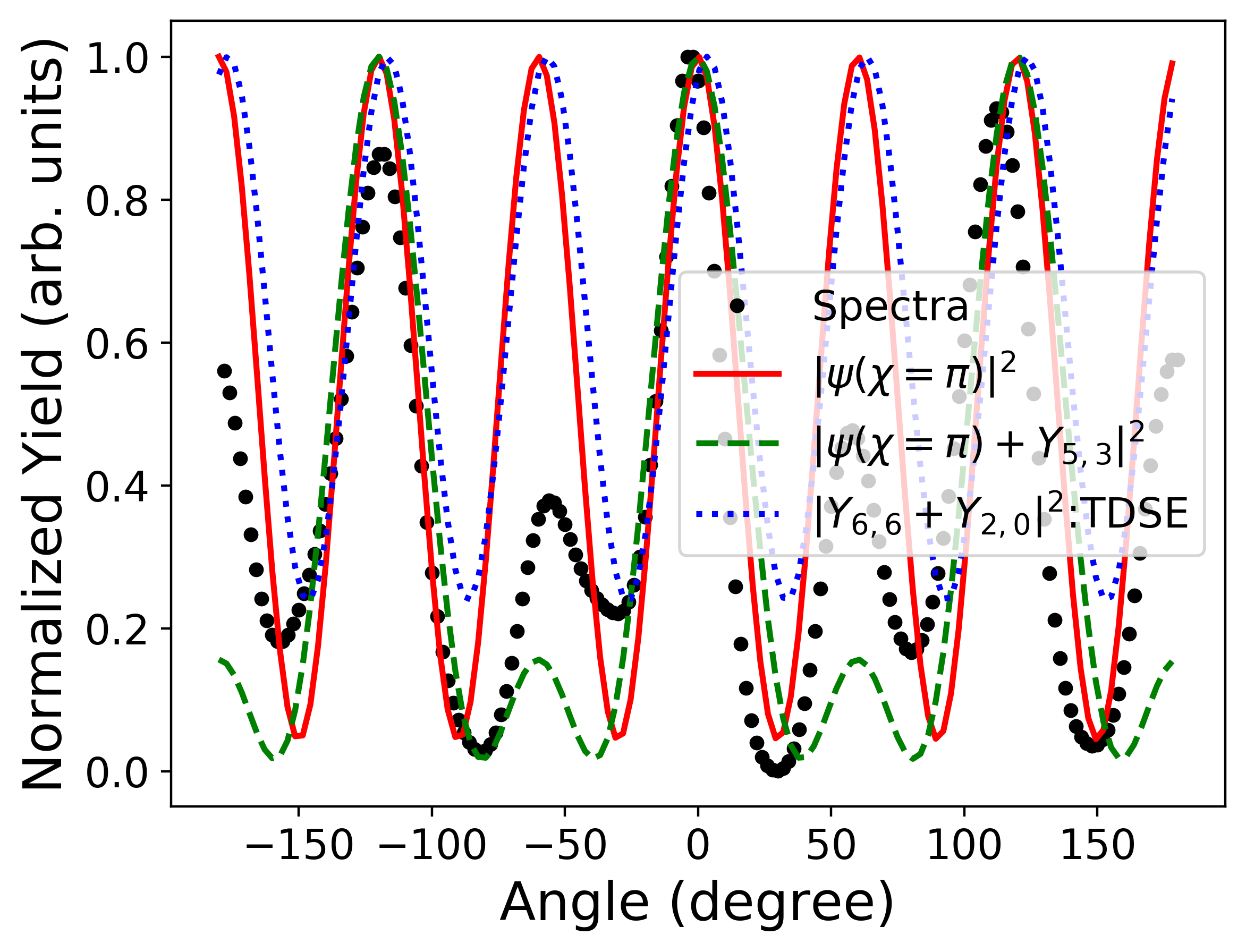}
	\caption{Comparison of the spectra at $|\vb{k}| =  0.03$ a.u. (dots) with the model of Eq.(\ref{eq:sph_harm_sum}), $|\psi(\theta = 0, \phi; \chi= \pi)|^2$ (red, solid) and including a second direct ionization channel $Y_{5,3}$ , i.e. $|\psi(\chi = \pi) + Y_{5,3}|^2$ (green, dashed). Coherent superposition of the states $Y_{6,6}$ and $Y_{2,0}$ with amplitude and phase from the solution of the TDSE (blue, dotted).}
	\label{fig:PES_vs_Ylm}
\end{figure}

The simplest approximation for the phase acquired is given by a classical oscillator, which changes its phase by $\chi = \pi$ when passing through a resonance. 
In Fig.\ref{fig:PES_vs_Ylm} the calculated spectra (black, dots) is compared with this interference model in the detector plane ($\theta = 0$) represented by $|\psi(\theta = 0, \phi; \chi= \pi)|^2$ is shown (red, solid) and reveal a very good agreement. In order to further verify this model the coherent superposition of the states represented by $Y_{6,6}$ and $Y_{2,0}$ with their corresponding amplitude and phase from the solution of the TDSE are shown in Fig.\ref{fig:PES_vs_Ylm} (blue, dotted) and the agreement emphasizes the validity of this model. The difference in magnitude of the different lobes can be understood from further contributions of other ionization channels.
Including the second most populated direct channel in the model, represented by $Y_{5,3}$ (green, dashed, Fig.\ref{fig:PES_vs_Ylm}), with no phase difference to the dominant direct channel, the relative intensity between two adjacent peaks approaches the numerical experiment.

Additional calculations for different pulse duration show that these striking structures already appear for pulses with a single cycle in the flat-top envelope region. Since ionization is extremely sensitive to the field strength and the resonant population of the intermediate states require absorption of several photons, such high-order processes from the leading and falling edge of the pulse are strongly suppressed and expected to give only small contributions. The fact that the low energy structure and the excited state population is already present for a pulse with a single cycle in the flat-top region, suggests that the interference pattern builds up within a single laser cycle. Thus we can attribute the appearance of these symmetric low energy structures to resonant bound-bound excitation and subsequent bound-continuum transitions from intermediate resonances interfering with direct bound-continuum pathways, within one laser cycle period of $\approx 2.7$ fs. This sub-cycle interference phenomena is supported by classical trajectory analysis \cite{eckart_subcycle}.

\begin{figure}
	\centering
	\includegraphics[width=7.0cm]{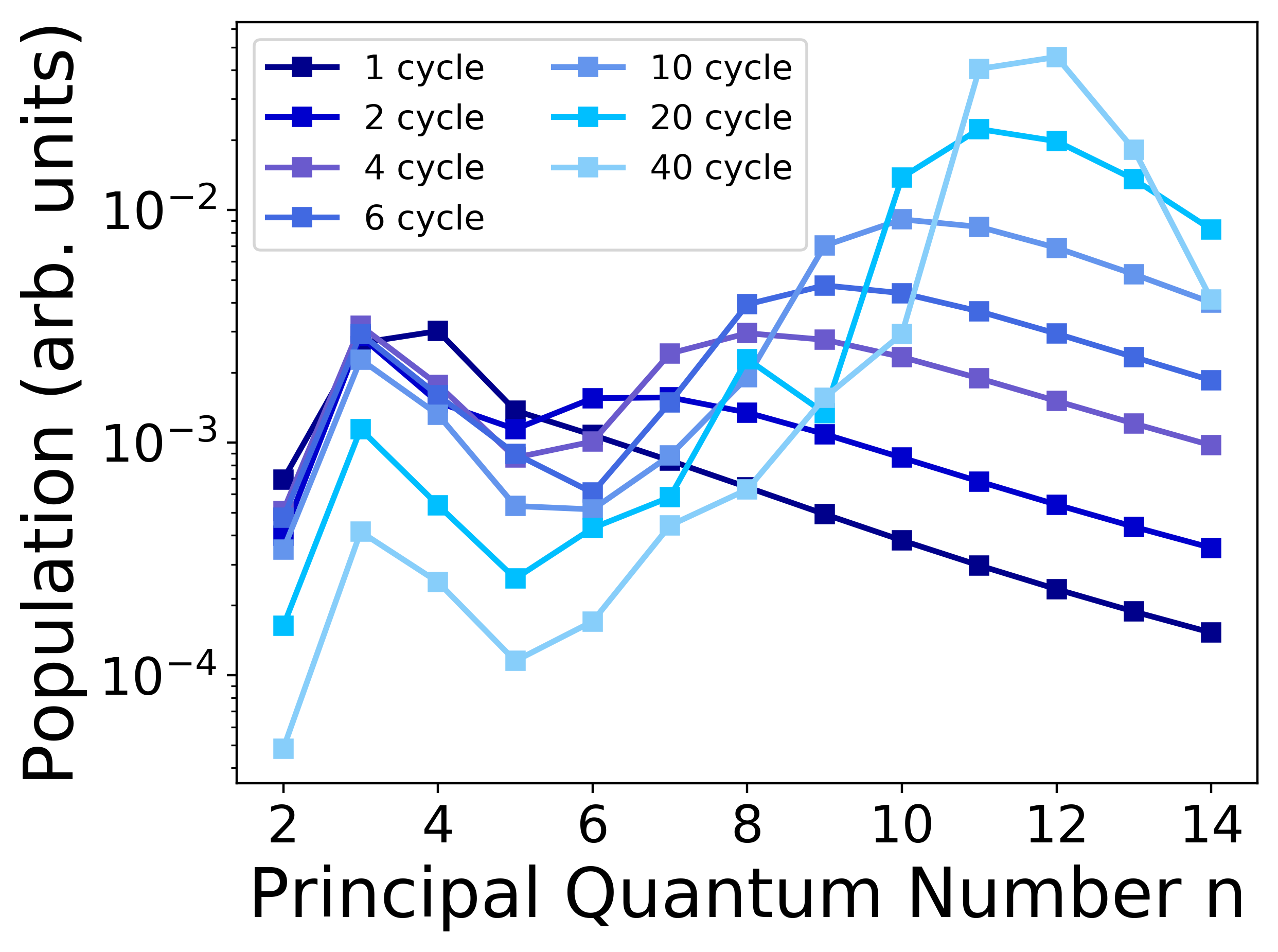}
	\caption{Distribution of the populated principle quantum number $n$ from the solution of the TDSE for different pulse duration (1-40 cycle) and a field strength of $F_0 = 0.03065$ atomic units. It shows the known dependence that the distribution shifts towards higher $n$ for longer pulses but exhibit a new feature which is independent of the pulse duration. Namely a peak at $n=3$ which is AC-Stark shifted into resonance. The initially populated $n=1$ ground state is not shown.}
	\label{fig:ndistribution}
\end{figure}

The above-mentioned resonant population enhancement is further reflected in the principal quantum number $n$ distribution after the end of the pulse, as it is shown in Fig.\ref{fig:ndistribution} for different pulse duration ($1-40$ flat top cycles, $\tau = 2.7-108$ fs) for a field strength of $F_0= 0.03065$ atomic units. The peak at $n=3$ is present for all pulse duration, and it is the dominant feature for the shortest pulses ($t_{flat}=1,2,4$ cycles). In agreement with the decrease of the $l = 1, m=-1$ population of the $n=3$ bound state (Fig.\ref{fig:population1} (a)) the total $n=3$ bound population in Fig.\ref{fig:ndistribution} starts to decrease at $t_{flat} = 10$ cycles. The virtually identical population for shorter pulses suggest that the population is trapped in the $l=1,m=-1$ bound state. In fact the behavior of the $n=3$ peak emphasize that the population is efficiently trapped in the $n=3$ manifold, or more precise in the $l=1, m=-1$ state, up to $\tau \approx 16.2$ fs (6 cycles), and gets depleted for longer pulses into higher Rydberg states or the continuum.

In contrast, for longer pulses, a secondary peak develops, which moves to higher principal quantum numbers with increasing pulse duration. In the case of longer pulses the electron can return more often to the vicinity of the core, and therefore has more chances to absorb an extra photon, as has been suggested for linear fields, and explained with the formation of Rydberg states in the semi-classical model of FTI \cite{eichmann_FTI,eilzer2014steering,ortmann2018controlling}. This leads to either population of Rydberg states with higher $n$ or to ionization, which would both cause the observed depletion of the $n=3$ resonance population for longer pulses and the emergence of the secondary peak. 

The different behavior of the two peaks, namely that the position of the $n=3$ peak is independent of the pulse duration whereas the second peak shifts to higher $n$ values for increasing pulse duration, suggest that two different mechanisms are responsible for their appearance, namely the resonant excitation mechanism and the semi-classical model supported by FTI. 

However, it should be noted, that the overall trapped population in the Rydberg states increases for increasing pulse duration. This increase of the excited state population can be explained in terms of channel-closing and the corresponding increase in the observation of excited state atoms for increasing field strength \cite{zimmermann2017unified}. Thus further mechanism affecting the observed Rydberg state population dynamics, such as channel-closing and envelope effects, can in principle not be excluded. Contributions from channel-closing could be explored by including the ground state AC-Stark shift, which would be responsible for a 12 photon channel-closing at this resonant field strength. However, envelope-effects do not seem to have a significant influence on the stability of the phase-relation between the ionization-pathways for the observed interference. Since for the field strength of this particular 11-photon resonance ($F_0 = 0.03065$ a.u.), a Keldysh parameter of $\gamma = 1.17$ suggests that the phase shifts due to the envelope are negligible in the regime where multi-photon processes dominate \cite{zimmermann2017unified}.

\section{\label{sec:conclusion}Conclusions}

In summary, we have presented photo-electron momentum distributions for the hydrogen atom exposed to bi-circular laser fields that exhibit a new structure at low energies. 
We discussed the origin of such structure in terms of interference between a resonant and a direct ionization pathway. The resonant pathway is due to AC-Stark shifted eigenstates, brought into resonance with the ground state dressed by the strong laser field. This field-strength dependent appearance of Freeman resonances in bi-circular laser fields and their influence on high harmonic generation has been shown \cite{jimenez2017time}, whereas clear signatures in the photo-electron momentum distribution has not been reported before. The detection of such Freeman resonances, to our knowledge had not been reported before in circularly polarized fields, monochromatic or otherwise.

In addition, we have analyzed the bound state population trapped in Rydberg states at the AC-Stark shifted resonance, and how it is related to the populations of continuum states. The principal quantum number distribution for various pulse duration exhibit a bimodal peak structure. The first peak at the $n=3$ resonance is virtually unaffected for pulse duration below 16 fs and already appears at a single cycle pulse of 2.7 fs. This further suggests that the creation of the resonant population is on the order, if not shorter than, the associated Kepler period of $2\pi n^3 \approx 4$ fs.

The fact that all the observed features attributed to the presence of the intermediate resonance, such as the low energy structure and the excited state population, are already present for a single cycle flat-top suggest that the interference pattern builds up within a single cycle. It further emphasize that such Freeman resonances can clearly and unambiguously be detected in the angular photo-electron momentum distribution for very short few-cycle, or even single-cycle, laser pulses.

In contrast, the detection of channel-closing and Freeman resonances in conventional energy resolved experiments is elusive, as, in order to resolve them in energy, one needs long pulses, hiding the sub-cycle nature of the excitation process. Thus, the unambiguous detection of Freeman resonances and its manifestation in the angular structure of the photo-electron spectrum provides a connection to the underlying Rydberg state manifold. This connection to the field-driven formation and dynamics of the Rydberg states, enabling detailed exploration and control of such dynamics, by controlling the relative parameters of the field.

Thus the interplay between channel-closing and the resonant excitation mechanism in tailored-intense fields, ranging from sub-cycle resonant population mechanism of Rydberg states to the excited state population trapping, seem to be promising in order to further understand and control the ultrafast dynamics of Rydberg state population in strong laser fields.

\section{\label{sec:acknowledgment}Acknowledgments}

The authors gratefully acknowledge stimulating discussions with Maria Richter, Olga Smirnova and Misha Ivanov in the early stages of the project.

\bibliographystyle{unsrt}
\bibliography{literatur}{}
\end{document}